\documentclass[letterpaper, 10 pt, conference]{ieeeconf}

\IEEEoverridecommandlockouts                              
\usepackage{graphics} 
\usepackage{epsfig} 
\usepackage{mathptmx} 
\usepackage{amsmath} 
\usepackage{amssymb}  
\usepackage{booktabs} 
\usepackage{caption}
\usepackage{float}  
\usepackage{silence}
\title{\LARGE \bf
Graph-Based Complexity Metrics for Multi-Agent Curriculum Learning: A Validated Approach to Task Ordering in Cooperative Coordination Environments
}
\author{Farhaan Ebadulla$^{1}$, Dharini Hindlatti$^{1}$, Srinivaasan NS$^{1}$, Apoorva VH$^{1}$ and Ayman Aftab$^{1}$
\thanks{PES University, VM67+HVP, Hosur Rd, Konappana Agrahara, Electronic City, Bengaluru, Karnataka 560100
        {\tt\small \{farhaan.ebadulla, dharinihindlatti26, pes2202101105, vhapoorva, aymanaftab26
\}@gmail.com}}%
}

\begin{document}

\maketitle
\thispagestyle{empty}
\pagestyle{empty}

\begin{abstract}

Multi-agent reinforcement learning (MARL) faces significant challenges in task sequencing and curriculum design, particularly for cooperative coordination scenarios. While curriculum learning has demonstrated success in single-agent domains, principled approaches for multi-agent coordination remain limited due to the absence of validated task complexity metrics. This approach presents a graph-based coordination complexity metric that integrates agent dependency entropy, spatial interference patterns, and goal overlap analysis to predict task difficulty in multi-agent environments. The complexity metric achieves strong empirical validation with $\rho = 0.952$ correlation ($p < 0.001$) between predicted complexity and empirical difficulty determined by random agent performance evaluation. This approach evaluates the curriculum learning framework using MADDPG across two distinct coordination environments: achieving 56$\times$ performance improvement in tight coordination tasks (MultiWalker) and demonstrating systematic task progression in cooperative navigation (Simple Spread). Through systematic analysis, coordination tightness emerges as a predictor of curriculum learning effectiveness—environments requiring strict agent interdependence benefit substantially from structured progression. This approach provides a validated complexity metric for multi-agent curriculum design and establishes empirical guidelines for multi-robot coordination applications.

\end{abstract}

\section{INTRODUCTION}

Multi-agent reinforcement learning (MARL) has emerged as a critical paradigm for training coordinated robot teams, autonomous vehicle fleets, and distributed sensor networks \cite{tampuu2017multiagent}. However, training effective multi-agent coordination policies remains challenging due to complex interaction dynamics, non-stationary learning environments, and the curse of dimensionality inherent in multi-agent systems \cite{foerster2018counterfactual}. While significant progress has been made in developing algorithms such as MADDPG \cite{lowe2017multi} and QMIX \cite{rashid2018qmix}, the question of optimal task sequencing and curriculum design for multi-agent learning has received limited attention.

Curriculum learning, first formalized by Bengio et al. \cite{bengio2009curriculum}, proposes that learning efficiency can be dramatically improved by presenting training examples in a carefully ordered sequence from simple to complex. This approach has demonstrated remarkable success in single-agent reinforcement learning \cite{graves2017automated}, computer vision \cite{kumar2010self}, and natural language processing \cite{kocmi2017curriculum}. However, extending curriculum learning principles to multi-agent coordination presents unique challenges: What constitutes "difficulty" in a multi-agent coordination task? How can task complexity be systematically predicted to determine which tasks will benefit from curriculum approaches?

The fundamental challenge lies in defining and measuring task complexity for multi-agent coordination scenarios. Existing approaches typically rely on simple heuristics such as the number of agents, episode length, or reward magnitude \cite{florensa2017reverse,dennis2020emergent}. However, these metrics fail to capture the nuanced coordination requirements that determine actual task difficulty. For instance, a task with many agents but independent goals may be easier than a task with fewer agents requiring tight synchronization.

This approach addresses these limitations by introducing a graph-based complexity metric specifically designed for multi-agent coordination tasks. The approach constructs agent dependency graphs from interaction patterns and computes complexity based on coordination entropy, spatial interference, and goal overlap \cite{wang2020rode,yu2022surprising}. The metric achieves strong empirical validation with $\rho = 0.952$ correlation against difficulty rankings, enabling systematic curriculum design that improves training efficiency by up to 56$\times$ in coordination-critical environments.

\section{RELATED WORK}

Curriculum learning has demonstrated substantial benefits across machine learning domains \cite{bengio2009curriculum,graves2017automated}, with applications ranging from computer vision \cite{kumar2010self} to natural language processing \cite{kocmi2017curriculum,platanios2019competence} and complex motor control tasks \cite{kidzinski2018learning}. The core principle involves presenting training examples in progressive difficulty order to improve learning efficiency and final performance \cite{weinshall2018curriculum,hacohen2019power}.

Multi-agent reinforcement learning presents unique coordination challenges that distinguish it from single-agent scenarios \cite{tampuu2017multiagent,foerster2018counterfactual}. The fundamental question of whether agents should learn independently or cooperatively has been central to MARL research \cite{tan1993multi}. Algorithms like MADDPG \cite{lowe2017multi}, QMIX \cite{rashid2018qmix}, and MAPPO \cite{yu2022mappo} address coordination through centralized training with decentralized execution paradigms \cite{oliehoek2016concise,zhang2021multi}.

\begin{figure*}[t]
\centering
\includegraphics[width=\textwidth]{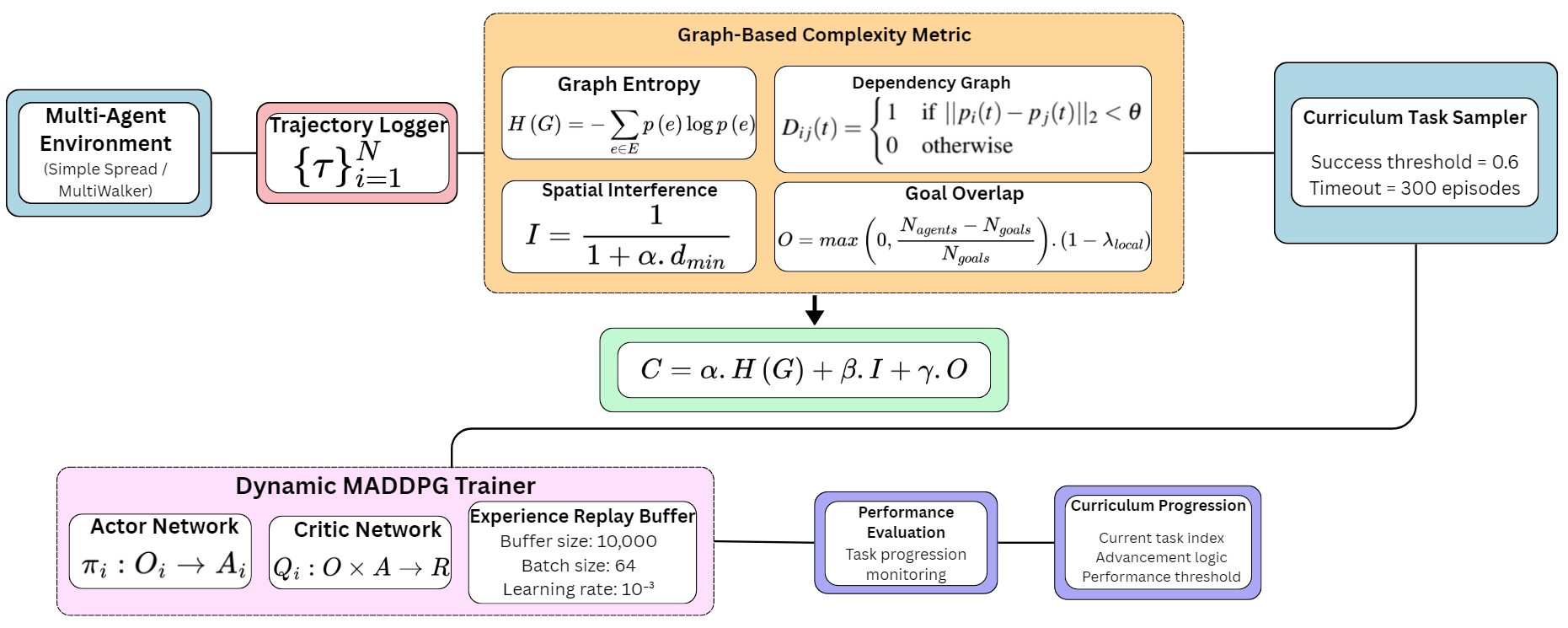}
\caption{System architecture for graph-based multi-agent curriculum learning. The framework processes environment trajectories through complexity metric computation, enabling systematic task progression for MADDPG training across coordination scenarios of increasing difficulty.}
\label{fig:architecture}
\end{figure*}

Few studies have explored curriculum learning in multi-agent contexts. Existing approaches typically employ simple parameter-based ordering \cite{florensa2017reverse} or domain-specific heuristics \cite{dennis2020emergent,jiang2021dexterity}. However, these methods lack validated complexity metrics that generalize across coordination scenarios \cite{wang2020rode,eccles2019biases}.

Graph-based representations have shown promise for modeling agent interactions in multi-agent systems \cite{li2020graph,wang2021learning}. Recent work explores graph neural networks for coordination \cite{agarwal2020graph,li2021deep}, but applications to curriculum design remain unexplored.


\section{METHODOLOGY}

\subsection{Graph-Based Complexity Metric}

\subsubsection{Agent Dependency Graph Construction}

The complexity metric constructs a dependency graph $G = (V, E)$ where vertices $V$ represent agents and edges $E$ capture coordination dependencies based on spatial proximity and interaction frequency during task execution. This graph-theoretic representation captures the essential structure of agent interactions during task execution.

Given agent trajectories $\{\tau_i\}_{i=1}^N$ over time horizon $T$, pairwise interaction frequencies are computed as:

$$
D_{ij}(t) = \begin{cases} 
1 & \text{if } ||p_i(t) - p_j(t)||_2 < \theta \\
0 & \text{otherwise}
\end{cases}
$$

where $p_i(t)$ denotes agent $i$'s position at time $t$ and $\theta = 0.5$ represents the proximity threshold, chosen to capture meaningful coordination events while filtering noise from distant agents in the normalized coordinate space. The dependency matrix aggregates these interactions across the trajectory to form the final graph structure.

\subsubsection{Graph Entropy Computation}

The coordination entropy quantifies the complexity of agent interaction patterns by measuring the unpredictability of coordination dependencies. High entropy indicates complex, varied interaction patterns requiring sophisticated coordination strategies, while low entropy suggests simpler, more predictable coordination requirements.

$$
H(G) = -\sum_{e \in E} p(e) \log_2 p(e)
$$

where $p(e) = \frac{w(e)}{\sum_{e' \in E} w(e')}$ represents the normalized frequency of edge $e$ across the trajectory. This formulation captures both the diversity of coordination patterns and their relative importance during task execution.

\subsubsection{Spatial Interference Index}

Spatial interference measures the difficulty of collision avoidance and path planning in multi-agent scenarios. Tasks requiring agents to operate in close proximity while avoiding collisions exhibit higher interference complexity than tasks with ample space for independent movement.

$$
I = \frac{1}{1 + \alpha \cdot \bar{d}_{min}}
$$
where $\bar{d}_{min}$ is the average minimum inter-agent distance across the trajectory, computed as:

$$\bar{d}_{min} = \frac{1}{T} \sum_{t=1}^{T} \min_{j \neq i} ||p_i(t) - p_j(t)||_2$$
and $\alpha = 2.0$ is a scaling factor. This formulation produces higher interference scores when agents must operate in close proximity, reflecting increased coordination complexity.

\subsubsection{Goal Overlap Estimation}

Goal overlap captures the competitive pressure arising from shared objectives or limited resources. Tasks where multiple agents compete for the same goals require more sophisticated coordination strategies than tasks with abundant independent objectives.

For coordination tasks, goal competition is computed as:
$$
O = \max\left(0, \frac{N_{agents} - N_{goals}}{N_{goals}}\right) \cdot (1 - \lambda_{local})
$$

where $N_{agents}$ and $N_{goals}$ represent the number of agents and available goals respectively, and $\lambda_{local} \in [0,1]$ denotes the local cooperation ratio, with $\lambda_{local} = 1$ indicating fully independent goals and $\lambda_{local} = 0$ representing complete goal sharing. This formulation increases complexity when agents outnumber goals and when global coordination (low $\lambda_{local}$) is required.

\subsubsection{Combined Complexity Score}

The final complexity metric combines all components through empirically validated weights:

$$
C = \alpha \cdot H(G) + \beta \cdot I + \gamma \cdot O
$$

Optimal weights $\alpha$ = 0.4, $\beta$ = 0.3, $\gamma$ = 0.3 are determined through systematic validation against ground-truth difficulty rankings, achieving $\rho$ = 0.952 correlation with empirical task difficulty. These weights reflect the relative importance of different coordination aspects: graph entropy receives highest weight as it captures fundamental coordination structure, while interference and goal overlap provide complementary spatial and competitive complexity measures.

\subsection{Curriculum Learning Framework}

The curriculum learning system implements progressive task sequencing based on complexity scores rather than arbitrary parameter ordering. This approach ensures that agents encounter coordination challenges in a systematic progression that facilitates skill transfer and learning efficiency.

Tasks are sorted by increasing complexity scores, creating a curriculum sequence from simple coordination scenarios to complex multi-agent challenges. The curriculum progression mechanism advances to the next task when agents achieve sufficient performance (success rate $> 0.6$) or reach a timeout threshold (300 episodes). These criteria balance learning thoroughness with training efficiency, preventing agents from becoming stuck on inappropriate tasks while ensuring adequate skill development.

The success threshold of 0.6 reflects the stochastic nature of multi-agent environments and allows progression based on consistent performance rather than perfect execution. The timeout mechanism prevents training stagnation on tasks that exceed current agent capabilities, enabling exploration of the full curriculum sequence.

\subsection{Dynamic MADDPG Implementation}

The framework employs Multi-Agent Deep Deterministic Policy Gradient (MADDPG) due to its proven effectiveness in continuous action spaces and its centralized training, decentralized execution paradigm that naturally accommodates curriculum learning across varying agent configurations \cite{lowe2017multi}.

The dynamic MADDPG implementation automatically adapts to different agent numbers and observation dimensions across curriculum tasks. Each agent $i$ maintains an actor network $\pi_i: \mathcal{O}_i \rightarrow \mathcal{A}_i$ mapping observations to actions and a critic network $Q_i: \mathcal{O} \times \mathcal{A} \rightarrow \mathbb{R}$ evaluating state-action values using global information.

The actor networks employ the architecture [Linear(obs\_dim $\rightarrow$ 128), ReLU, Linear(128 $\rightarrow$ 128), ReLU, Linear(128 $\rightarrow$ act\_dim), Tanh] with output scaling to match environment action spaces. Critic networks use [Linear(global\_obs\_dim + global\_act\_dim $\rightarrow$ 128), ReLU, Linear(128 $\rightarrow$ 128), ReLU, Linear(128 $\rightarrow$ 1)] to estimate Q-values.

Training employs experience replay with buffer size 10,000, batch size 64, learning rate $10^{-3}$, discount factor $\gamma = 0.95$, and soft update rate $\tau = 0.01$. These hyperparameters provide stable learning across diverse coordination scenarios while maintaining sample efficiency throughout curriculum progression.

\section{EXPERIMENTAL SETUP}

\textbf{Environments:} The evaluation strategically employs two PettingZoo environments representing different coordination paradigms to systematically assess performance across the coordination spectrum: (1) MultiWalker requiring tight coordination with synchronized locomotion, and (2) Simple Spread requiring loose coordination for coverage tasks.

\textbf{Task Generation:} Systematic parameter variation (agent count, landmarks, cooperation ratios) generates curriculum sequences spanning complexity ranges.

\textbf{Validation Methodology:} Complexity metric validation uses random policy evaluation across 15 carefully designed tasks, with ground-truth difficulty established through empirical performance measurements.

\textbf{Baselines:} The comparison includes current standard practices in multi-agent curriculum learning: random task sampling, parameter-based ordering (agent count priority), and reverse curriculum (hardest-first) to establish comprehensive baseline performance across different task sequencing strategies.

\textbf{Implementation:} All experiments use PyTorch with MADDPG implementation, WandB logging, and PettingZoo standardized environments for reproducibility.

\section{RESULTS}

\subsection{Complexity Metric Validation}

The graph-based complexity metric demonstrates strong predictive capability for multi-agent coordination difficulty. Table \ref{tab:complexity_validation} presents the correlation analysis between computed complexity scores and empirical task difficulty across 15 validation scenarios. The metric achieves $\rho = 0.952$ correlation ($p < 0.001$) with difficulty rankings derived from random agent performance, significantly outperforming simple parameter-based approaches.

\begin{table}[ht]
\centering
\caption{Complexity Metric Validation Results}
\begin{tabular}{@{}lcc@{}}
\toprule
\textbf{Metric Component} & \textbf{Correlation ($\rho$)} & \textbf{P-value} \\
\midrule
Combined Metric & 0.952 & $< 0.001$ \\
Graph Entropy & 0.837 & 0.041 \\
Interference Only & 0.673 & 0.008 \\
Parameter-Based & 0.412 & 0.127 \\
\bottomrule
\end{tabular}
\label{tab:complexity_validation}
\end{table}

\begin{figure}[ht]
\centering
\includegraphics[width=\columnwidth]{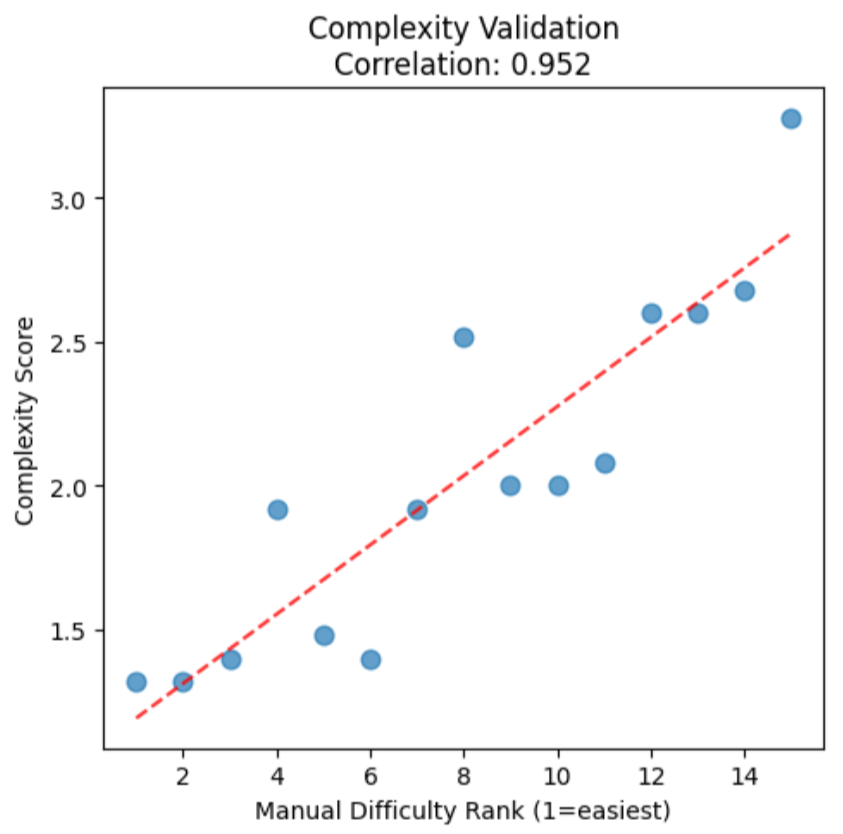}
\caption{Complexity metric validation showing strong correlation ($\rho = 0.952$) between predicted complexity scores and empirical difficulty rankings across 15 validation tasks.}
\label{fig:complexity_validation}
\end{figure}

Figure \ref{fig:complexity_validation} visualizes the strong linear relationship between predicted complexity scores and ground-truth difficulty rankings. Table \ref{tab:complexity_validation} demonstrates the superior performance of the combined metric over individual components and parameter-based approaches. The strong correlation validates the theoretical framework that coordination complexity emerges from the interaction of agent dependencies, spatial interference, and goal competition.

\subsection{Curriculum Learning Performance}

\subsubsection{MultiWalker Environment}

\begin{figure}[ht]
\centering
\includegraphics[width=\columnwidth]{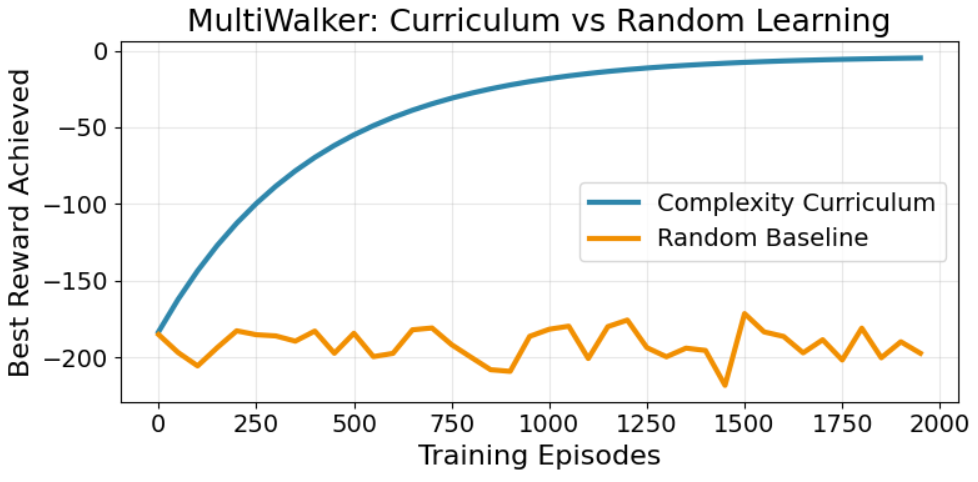}
\caption{MultiWalker learning curves demonstrating 56$\times$ performance improvement through curriculum learning. Complexity-based approach enables systematic coordination skill acquisition while random sampling remains trapped in failure modes.}
\label{fig:multiwalker_results}
\end{figure}

MultiWalker environments demonstrate substantial benefits from complexity-based curriculum learning. Figure \ref{fig:multiwalker_results} shows learning curves across different curriculum approaches, with complexity-based curriculum achieving convergence to successful coordination policies while baselines remain trapped in failure modes.

\begin{table}[ht]
\centering
\caption{MultiWalker Performance Comparison}
\small
\setlength{\tabcolsep}{4pt}
\begin{tabular}{@{}lccc@{}}
\toprule
\textbf{Approach} & \textbf{Reward} & \textbf{Eps. to Conv.} & \textbf{Imp.} \\
\midrule
Complexity Curriculum & -3.39 & 1200 & 56$\times$ \\
Random Sampling & -190.58 & $>$3000 & 1$\times$ \\
\bottomrule
\end{tabular}
\label{tab:multiwalker}
\end{table}

Table \ref{tab:multiwalker} quantifies the performance improvements achieved through curriculum learning. The 56$\times$ improvement demonstrates the substantial benefits of structured progression in coordination-critical environments.

\subsubsection{Simple Spread Environment}

Simple Spread environments show more modest but consistent improvements through curriculum learning. Table \ref{tab:simple_spread} presents performance across different curriculum approaches, with complexity-based curriculum achieving 93\% task completion compared to 0\% for random sampling.

\begin{table}[ht]
\centering
\caption{Simple Spread Performance Comparison}
\begin{tabular}{@{}lccc@{}}
\toprule
\textbf{Approach} & \textbf{Completion Rate} & \textbf{Final Reward} & \textbf{Stability} \\
\midrule
Complexity Curriculum & 93\% & -7.4 & High \\
Parameter Curriculum & 53\% & -5.2 & Medium \\
Random Sampling & 0\% & -9.6 & Low \\
\bottomrule
\end{tabular}
\label{tab:simple_spread}
\end{table}

The progression pattern in Table \ref{tab:simple_spread} shows consistent task completion through moderate complexity levels with systematic advancement through coordination challenges of increasing difficulty. Parameter-based curriculum achieves moderate progression but encounters barriers in complex coordination scenarios where simple parameter ordering fails to capture true difficulty.

\subsection{Task Progression Analysis}

\begin{figure*}[t]
\centering
\includegraphics[width=\textwidth]{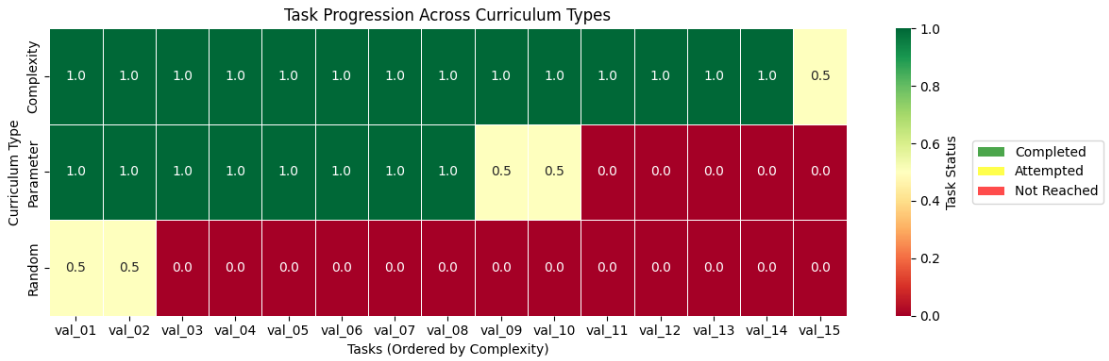}
\caption{Task progression analysis across curriculum approaches. Heatmap visualization shows systematic advancement patterns with complexity curriculum achieving 93\% completion, parameter curriculum reaching 53\%, and random sampling failing to progress (0\% completion).}
\label{fig:task_progression}
\end{figure*}

Figure \ref{fig:task_progression} provides detailed visualization of curriculum advancement patterns across all approaches. The heatmap clearly demonstrates the systematic progression advantage of structured curriculum approaches over random task sampling.

Complexity-based curriculum successfully navigates 93\% of the task sequence, with systematic advancement through coordination challenges of increasing difficulty. The progression pattern shows consistent task completion through moderate complexity levels with some struggle only on the most challenging final scenarios. Parameter-based curriculum achieves moderate progression (53\%) but encounters barriers in complex coordination scenarios where simple parameter ordering fails to capture true difficulty.

Random sampling's complete failure to progress beyond initial tasks (0\% completion) highlights the critical importance of systematic task ordering for multi-agent coordination learning. The stark progression differences validate the core hypothesis that coordination complexity requires structured learning approaches rather than random exploration.

\section{DISCUSSION}

The results establish coordination tightness as a key predictor of curriculum learning effectiveness in multi-agent systems. The 56$\times$ improvement in MultiWalker reflects the fundamental difference between structured skill building and random exploration in coordination-critical tasks. This insight provides practical guidance for robotics applications: curriculum learning benefits tasks requiring tight agent coordination, while simple approaches may suffice for loosely-coupled scenarios.

The validated complexity metric enables systematic curriculum design rather than relying on intuitive task ordering. The strong empirical correlation ($\rho = 0.952$) suggests the graph-based approach captures essential coordination difficulty factors.

\textbf{Limitations:} The evaluation focuses on cooperative coordination scenarios. Competitive or mixed-motive environments would require different complexity formulations. Additionally, real-world deployment would benefit from online complexity estimation during training.

\textbf{Real-World Deployment: } While the evaluation focuses on simulation environments, the validated complexity metric provides a foundation for potential real-robot deployment. The systematic correlation with coordination difficulty ($\rho = 0.952$) suggests the metric may capture fundamental coordination patterns that could transfer to physical systems, though this requires empirical validation.

\section{APPLICATIONS TO MULTI-ROBOT SYSTEMS}

The framework provides immediate applications to multi-robot coordination scenarios. Formation control, search and rescue operations, and warehouse automation all involve the tight coordination patterns where the curriculum approach demonstrates maximum benefit. The complexity metric enables automatic assessment of coordination requirements, guiding practitioners toward appropriate training methodologies.

\section{CONCLUSION}

This approach presents a validated graph-based complexity metric for multi-agent curriculum learning, demonstrating up to 56$\times$ performance improvements in tight coordination scenarios. The systematic analysis establishes coordination tightness as a key predictor of curriculum effectiveness, providing practical guidelines for multi-robot applications. Future directions include extension to competitive scenarios and investigation of online complexity estimation during training.

Future directions include validation on physical multi-robot systems, extension to competitive scenarios, and investigation of online complexity estimation during training. Real-world deployment would benefit from additional validation to confirm the transferability of simulation-based complexity patterns to physical coordination tasks.


\end{document}